\documentclass[aps,preprint,letterpaper,groupeaddress,showpacs]{revtex4}
\usepackage{epsfig}                       \usepackage{graphicx}
\usepackage{psfrag}                       \usepackage{color}
\usepackage{psfig}                        \usepackage{amsbsy}
\usepackage{amssymb}

\begin{document}
\bibliographystyle{apsrev}
\title{First Principles Study of Work Functions of Single Wall Carbon Nanotubes}
\author{Bin Shan}
\affiliation{Department of Applied Physics, Stanford University, Stanford CA, USA, 94305-4040}
\author{Kyeongjae Cho}
\email{kjcho@stanford.edu}
\affiliation{Department of Mechanical Engineering, Stanford University, Stanford CA, USA, 94305-4040}
\date{\today}

\begin{abstract}
We performed first principles calculations on work functions of single wall carbon nanotubes (SWNTs), which can be divided into two classes according to tube diameter ($D$). For class $I$ tubes ($D>1 nm$), work functions lie within a narrow distribution ($\sim 0.1 eV$) and show no significant chirality or diameter dependence. For class $II$ tubes ($D<1 nm$), work functions show substantial changes, with armchair tubes decrease monotonously with diameter, while zigzag tubes show opposite trend. Surface dipoles and hybridization effects are shown to be responsible for the observed work function change.
\end{abstract}
\maketitle
Carbon nanotubes (CNTs)\cite{IIJIMA} have attracted considerable attention as a promising candidate material for future nano-scale electronics. It boasts reliable structural strength, mature synthesis methods, and unique electronic structure that is closely related to its chirality. Nanotube with indices ($n$,$m$) can be either metallic or semiconducting depending on whether $n-m$ is a multiple of three. Besides bandgap, absolute
potential of the fermi level, namely the work function, is another important physical quantity of interest. Nanotube work function is a critical quantity in understanding  field emission properties from SWNTs\cite{GZhou}. It dictates the direction of charge transfer and influences the selectivity of molecules as chemical sensors\cite{JKong,Shu}. It affects band lineup at the CNT/metal contact\cite{BSHAN} and has considerable impact on device performance\cite{Heinze}. From device integration perspective, it is desirable to have all the nanotubes acting as channel materials with the same work function. All these indicate the importance of a thorough understanding of nanotube work functions.

However, there have only been few theoretical studies on the work functions of SWNTs. Simple $\pi$-orbital tight binding, which serves as a good starting point for understanding nanotube electronic structure, would predict no work function change at all. Zhao $et$ $al$\cite{Zhao} studied work functions of CNTs based on the molecular orbital calculations, indicating higher work function of semiconducting nanotubes and considerable diameter dependence over all diameter range. Some other $ab$ $inito$ calculations\cite{BLASE,Park,Mazzon} have investigated curvature effects and their effects on the band gap. Less is clear about their implications on work functions.  On the experimental side, it has recently become possible to measure work functions of individual SWNTs\cite{Okazaki,Suzuki}. However, there is so far no consensus among different experiments. Resonant confocal micro Raman spectroscopy\cite{Okazaki} reveals considerable work function difference between metallic and semiconducting nanotubes while photoemission electron microscopy study\cite{Suzuki} shows work functions of various nanotubes is generally insensitive to either diameter or chirality. In this paper, we employed first principles plane-wave pseudo potential calculations and systematically studied work functions of ($n$,$0$) and ($n$,$n$) nanotubes.  The nanotubes we studied range from ultra small tubes $\sim 0.4 nm$ to conventional size tubes $\sim 2.5 nm$ usually grown by chemical vapor deposition (CVD). Work function study of ultra small tubes is motivated by recent experimental fabrication and observation of nanotubes with diameters down to $\sim 0.4 nm$\cite{ZMLi,Qin,Wang,Peng}. From a systematic analysis of the calculation results, work function behavior of nanotubes can be divided into two classes. Work functions of class $I$ tubes fall in a narrow distribution $\sim0.1 eV$. For class $II$ tubes, work functions increase or decrease significantly depending on whether it is armchair or zigzag tube.

Workfunction is defined as $WF=\phi-E_f$, where $\phi$ is the vacuum level and $E_f$ is the Fermi level of the system. In the case of semiconducting nanotubes, we place $E_f$ at the midgap, which is consistent with experimental measurements\cite{Okazaki,Suzuki}. The work function calculation is done using Vienna Ab-initio Simulation Package (VASP)\cite{VASP}. To determine the work function, the nanotube is put into a supercell of 35$\AA \times 35\AA$ in the cross sectional direction. This corresponds to a nearest inter-tube distance of $>10\AA$ for the largest tube under study. Kohn-Sham single-electron wavefunctions are expanded by 921984 plane waves with an energy cut off of 21.06 Ryd. 30 k-points are used along the one dimensional Brillouin zone which are tested to give good convergence. All the atoms in the unit cell are fully relaxed until the force on each atom is less than $0.02 eV/\AA$. For class $I$ tubes, there is little difference between the optimized geometry and rolling up of a graphene sheet. For class $II$ tubes, the main effect of relaxation is the expansion of the tube diameter  by a few percent\cite{Machon}. The vacuum level $\phi$ is determined from the average potential in the vacuum region where it approaches a constant. Local density approximation (LDA) are used throughout for the exchange-correlation function. It is known that LDA utilized with Kohn-Sham scheme\cite{Kohn} tends to underestimate the bandgap, which in our case, results in an over estimate of the work function for semiconducting tubes. For an estimate of the error that may arise from band gap underestimate for semiconducting nanotubes in class $I$, we compare our calculated LDA bandgap with Tight binding(TB) and GGA calculations. Table I lists the bandgap for selected CNTs. It can been seen that for reasonably large diameter tubes, the band gaps derived from different theoretical models differ by less than $0.1eV$. This translates to an error $\sim$0.05eV in work function and will not affect our conclusions. Also, ($n$,$0$) tubes with $n<6$ become metallic due to strong curvature effects and we expect LDA to give a reasonably accurate description of their electronic structures. The largest possible error in work function estimate then occurs for (7,0) and (8,0) semiconducting tubes, which belong to class $II$ tubes. One way to correct the error is to use Hedin's GW method\cite{HEDIN} which has been shown to give accurate results for a wide range of materials\cite{Rohlfing}. GW calculations\cite{Satta} confirm the LDA result that (6,0) CNT is metallic and (7,0) is semiconducting (Table I). Thus corrections from GW method will only shift work function values for (7,0) and (8,0) tubes without altering the overall trend of work function change.

\begin{figure}
    \includegraphics[width=3.0in]{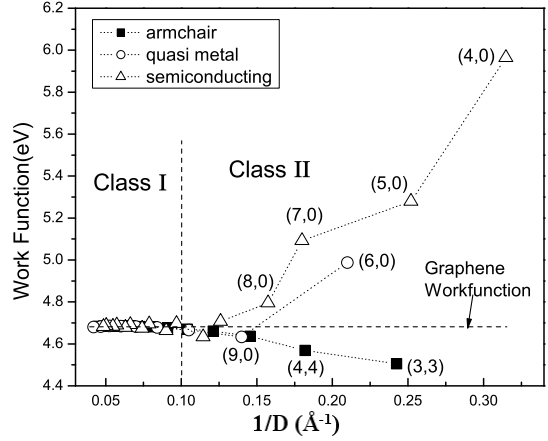}
    \caption
    {Work functions of ($n$,$n$) and ($n$,$0$) nanotubes of different diameter. For class $I$ tubes, nanotube work functions are very close to the work function of a graphene sheet $\sim4.66eV$. For class $II$ tubes, work functions of ($n$,$0$) tubes increase dramatically, while those of ($n$,$n$) tubes show decrease in work function. (6,0),(5,0) and (4,0) tubes are metallic due to $\sigma^*-\pi^*$ hybridization. }
    \label{WorkFunction}
    \end{figure}

Fig.\ref{WorkFunction} summarizes the work functions of ($n$,$0$) and ($n$,$n$) nanotubes with diameters ranging from $0.32 nm$ to $2.5 nm$, plotted against the inverse tube diameter. The dotted horizontal line in the graph indicates the calculated work function for a graphene sheet of $4.66 eV$. For class $I$ nanotubes, work functions of both semiconducting and metallic tubes fall within a very narrow distribution around $4.66 eV$, just as would be predicted from $\pi$-orbital tight binding model. More detailed analysis shows the work function decreases slightly with diameter. If work functions of class $I$ tubes are fitted to a curve of the form $y=a/D+4.66 eV$, then $a=-0.2\pm0.08\hspace{0.2cm} eV\cdot\AA$. The slope of the decrease is so small that for practical purposes, work functions of class $I$ tubes can be well approximated by the graphene work function. We can conclude that for class $I$ nanotubes, there is no significant work function difference within LDA framework. Since the tubes used in both experiments\cite{Okazaki,Suzuki} are larger than $1nm$ in diameter, our calculation is more consistent with Suzuki's\cite{Suzuki} finding that there is no structural dependence on the work functions of CNTs. The discrepancy between our prediction and that of Zhao's\cite{Zhao} regarding work functions of semiconducting nanotubes largely lies in the difference of work function definition, where they approximated the work function as $WF=-E_{HOMO}$ for semiconducting tubes. Thus the larger work functions of semiconducting tubes and inverse diameter dependence in Zhao's study is primarily due to the changes in band gap. Comparing with their results $WF=-E_f$ for metallic nanotubes, we predict an even weaker dependence on diameter for class $I$ tubes by including contribution from vacuum level. Besides these differences, our calculation for class $I$ tubes is generally consistent with the results from molecular orbital calculations. However, significant changes in work functions can be observed for class $II$ tubes. As can be seen from Fig. \ref{WorkFunction}, work functions of class $II$ nanotubes split distinctively into two groups. Work functions of ($n$,$n$) tubes monotonously decreases with diameter, while work functions of ($n$,$0$) tubes show a dramatic increase. The magnitude of the work function change is comparable to nanotube bandgap and thus plays an important role in determining the band lineup\cite{Leonard}. Its consequences can be experimentally observed via transport measurements\cite{Heinze,Javey}. For the smallest tube we studied, (4,0) tube has an ultra high work function of $5.95eV$.

\begin{figure}
    \includegraphics[width=3.0in]{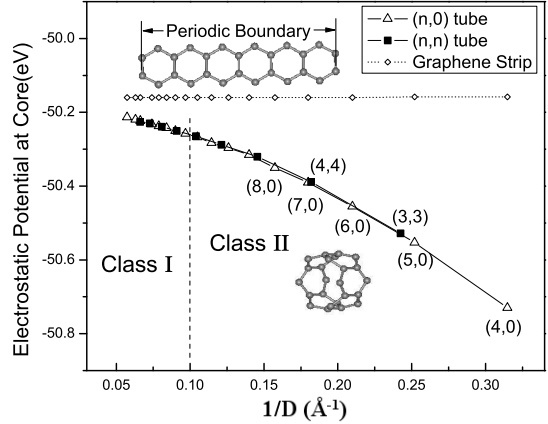}
    \caption
    {As tube diameter decreases, there is a lowering of the average potential at carbon atomic core with respect to vacuum which is due to asymmetric charge distribution inside and outside the tube. This effect is more pronounced for class $II$ tubes where a slight change in diameter leads to a large reduction in potential energy. The graphene sheet with corresponding boundary condition does not show such reduction.}
    \label{POT}
    \end{figure}

Work function increase(decrease) can be induced by either an enhanced(reduced) surface dipole, or a lowering(rising) of its intrinsic bulk Fermi energy. To gain a quantitative understanding from these two contributions, we rewrite work function as
\begin{equation}
WF=\phi-E_f=[\phi-V_{ref}]-[E_f-V_{ref}]
\label{eqn1}
\end{equation}
where $V_{ref}$ represents average electrostatic potential at carbon atomic core. In this expression, the first term is the potential with respect to the vacuum level (differ by a minus sign), whose change reflects a change in surface dipole contribution as we outline below. The second term represents the relative position of the Fermi energy above the average potential, which is related to the material's intrinsic bulk electronic structure. As shown in Fig.\ref{POT}, the average potential at the carbon atom monotonously decreases with decreasing diameter. By comparing with the average potential in a graphene sheet of corresponding periodic boundary condition, it can be concluded that the lowering of the potential is not due to quantization along the circumference in a nanotube, but rather due to curvature effect. For large diameter tubes, the potential inside and outside the tube are almost equal, indicating symmetric charge distribution. As the tube becomes smaller, the potential inside the tube falls below the vacuum level outside, indicating asymmetric charge distribution, with more charge distributed outside. More asymmetric charge distribution results in an enhanced surface dipole and lowers $V_{ref}$ with respect to vacuum level. The potential lowering due to surface dipole is insignificant for class $I$ tubes, but it can be up to $0.5 eV$ for class II tubes. The first term in Eqn.(\ref{eqn1}) alone, however, does not dictate the trend of work function change. Accompanying the lowering of the average potential is the tighter confinement of wave function along the circumferential direction, and the electron fermi level is generally expected to rise with respect to $V_{ref}$, which counteracts the effect of work function increase due to potential lowering. In fact, for class $I$ nanotubes, the second term increases faster than the first term, resulting in a slightly decreasing work function as the tube diameter decreases.

\begin{figure}
    \includegraphics[width=2in]{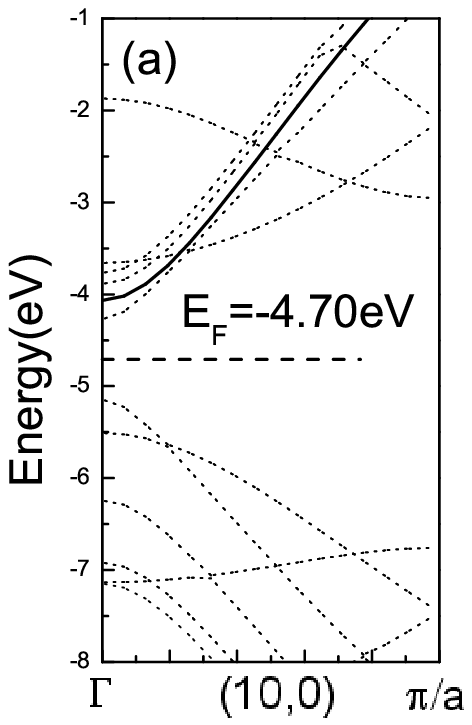}
    \includegraphics[width=2in]{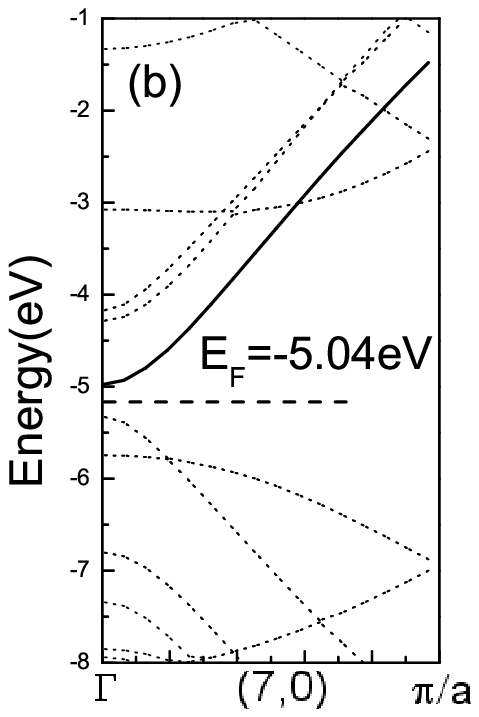}
    \includegraphics[width=2in]{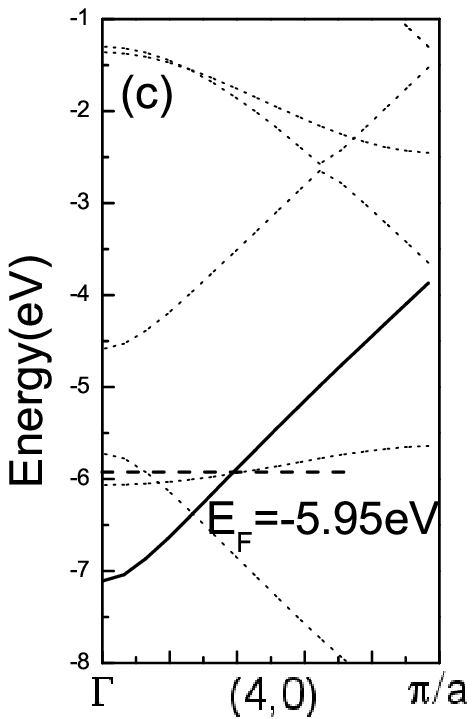}
    \caption
    {Band structures of (a)(10,0) (b) (7,0) (c) (4,0) CNTs respectively. here $a=3a_0$, where $a_0=1.44\AA$ is the lattice constant of graphene. The singly degenerate state (solid line) lies above the conduction band bottom (a), in the bandgap (b) and below the valence band top (c) respectively. Accompanying the lowering down of the singly degenerate state is the lowering of the Fermi level with respect to the vacuum. }
    \label{Bands}
    \end{figure}

The above picture holds true for class $I$ tubes. However, the second term $E_f-V_{ref}$ decreases rapidly for zigzag nanotubes smaller than (10,0). Combined with the potential lowering from the first term, this makes zigzag tubes' work functions increase dramatically. This is exactly where a singly degenerate state in the conduction band moves down below the conduction band bottom due to $\sigma^*-\pi^*$ hybridization\cite{BLASE}. In Fig. \ref{Bands} , we show band structures of (10,0), (7,0) and (4,0) CNTs respectively. In (10,0) CNT, the conduction band bottom is doubly degenerate as is predicted by simple $\pi$-orbital tight binding model. The singly degenerate state (indicated by a solid line in the figure), which originates from the band that crosses $\Gamma$ in the Brillouin zone lies approximately $1eV$ above the conduction band bottom. Upon increasing curvature, this singly degenerate bands is significantly lowered. In (7,0) CNT, it is below the $\pi^*$ conduction band bottom and lies within the bandgap, whereas for (4,0), (5,0) and (6,0) CNTs, it is even lowered below the $\pi$ valence band top, transforming them into metallic tubes. The lowering of this singly degenerate band due to curvature effects has been observed in many previous theoretical studies\cite{BLASE,Park,Mazzon}, however, its implication on work function change has not been examined. Most of the earlier works have focused on its effect on band gap change, and the energy levels are studied with respect to the fermi level, which is usually set to zero.

\begin{figure}
    \includegraphics[width=3in]{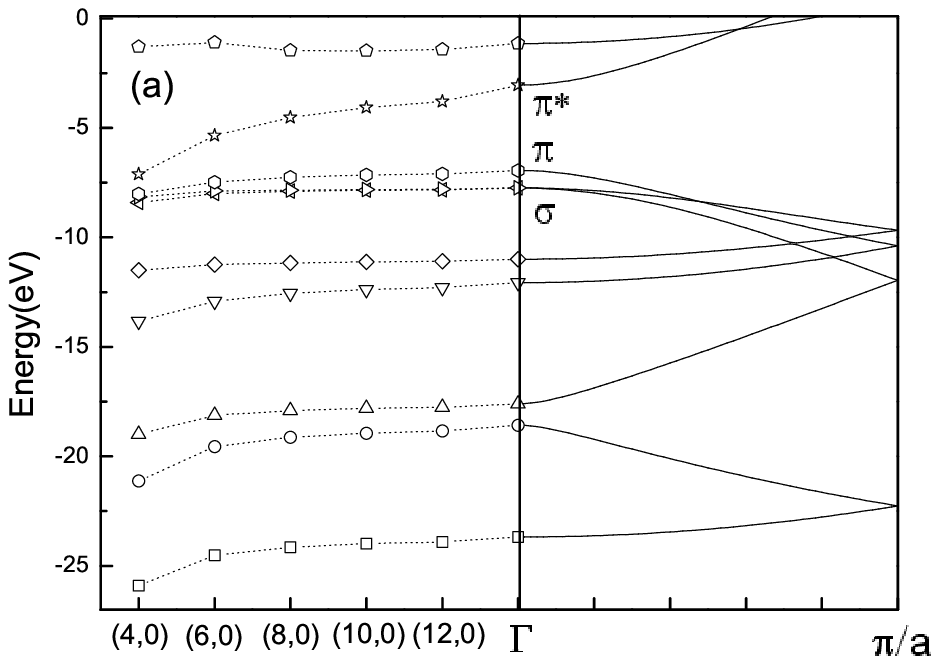}
    \includegraphics[width=3in]{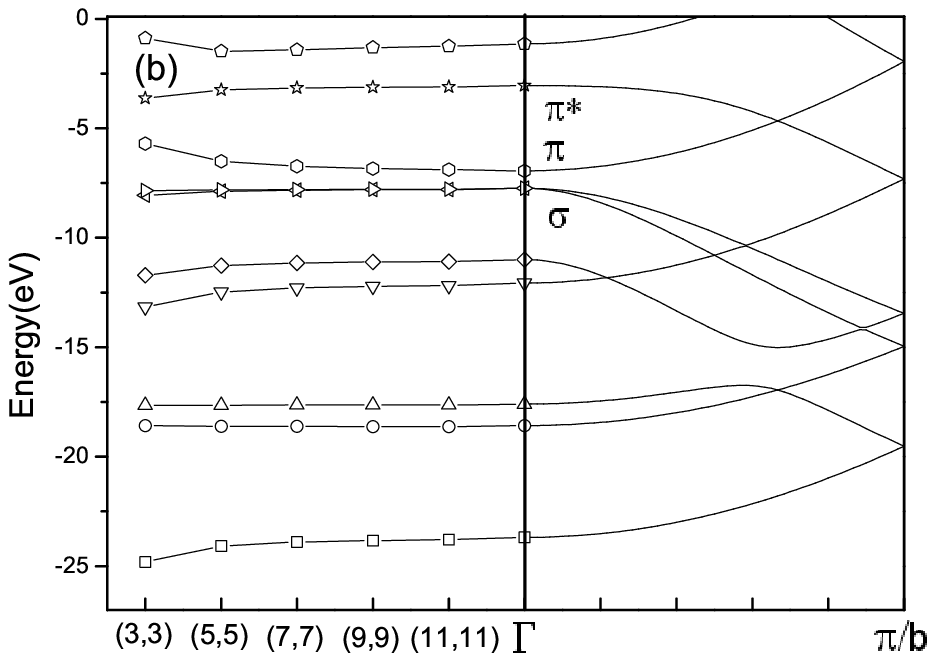}
    \caption
    {Evolution of the singly degenerate energy levels at $\Gamma$ point for (a) ($n$,$0$) and (b) ($n$,$n$) CNTs. On the right hand side of both graphs is the graphene band structure folded at the middle of Brillouin zone, where $a=3a_0$ and $b=\sqrt{3}a_0$. On the left hand side of both graphs is the absolute energy values of singly degenerate states in the nanotube, which is derived from the band that crosses $\Gamma$ point in graphene in the zone folding model.}
    \label{Elevel}
    \end{figure}

In the following, we show that the lowering of this singly degenerate band has important effects on work function. Fig. \ref{Elevel}a shows the graphene band structure along $\Gamma-M$ line folded at the middle of Brillouin zone and energy levels of the corresponding singly degenerate bands at $\Gamma$ point in a ($n$,$0$) nanotube. All the energy levels are drawn in absolute scale with respect to vacuum (which is set to be zero). The energies of these bands at $\Gamma$ point show a downward shift compared to that of graphene, indicating the lowering of the potential inside the tube (Fig. \ref{POT}). The top of the singly degenerate $\pi$ band is lowered $\sim1.0eV$ in going from graphene to (4,0) CNT, and $\sigma$ band is lowered by $\sim 0.5eV$, roughly in accordance with the lowering in $V_{ref}$. The effect of $\sigma^*-\pi^*$ hybridization induced by curvature is to bring down the singly degenerate $\pi^*$ band with respect to the already lowered $\sigma$ and $\pi$ bands. Before it moves below the doubly degenerate conduction band bottom, the work function is largely determined by the doubly degenerate valence band top and conduction band bottom. Once this singly degenerate states is down shifted below the conduction band bottom, as is the case for (7,0)and (8,0) CNT, there is an accompanying increase in work function due to the lowering of the conduction band edge. For zigzag tubes smaller than (6,0), this singly degenerate state crosses the valence band, pinning the fermi level below the valence band top. The concerted effects of potential lowering and moving down of the singly degenerate state makes the (4,0) CNT a material with extremely high work function. Fig. \ref{Elevel}b shows the same plot for ($n$,$n$) tubes. $\sigma$ band top is also lowered by similar amount as $V_{ref}$, but due to weaker $\sigma^*-\pi^*$ interaction in armchair tubes\cite{Reich}, there is no appreciable lowering of the singly degenerate $\pi^*$ state. In fact, more pronounced effect is the upshift of the bonding $\pi$ orbital due to enhanced $\sigma-\pi$ interaction\cite{Park} which results in an decrease in work function.

In conclusion, we have studied work functions of intrinsic SWNTs with experimentally observable diameter ranging from $\sim0.4nm$ to $\sim2.5nm$. Work functions of class $I$ tubes show little dependence on diameter or chirality. For class $II$ tubes, work functions show significant differences for ($n$,$0$) and ($n$,$n$) tubes. $\sigma^*-\pi^*$ interaction which pushes a singly degenerate state to cross the valence band is the main reason for the dramatic increase of work function observed in ($n$,$0$) tubes.

\begin{acknowledgments}
    This work is supported by NSF grant on Network for Computational Nanotechnology (NCN). The authors wish to thank Avik Ghosh and Diego Kienle for helpful suggestions.
\end{acknowledgments}

\begin{table}
\caption{Band gaps of selected CNTs}
\begin{tabular}{lcccc}
\hline\hline
Chirality   &      TB*    & GW\cite{Satta} &  GGA\cite{Sun}   &   Our Data(LDA)   \\
\hline
(6,0)       &             0.14     &  metal   &    metal  &   metal \\
\hline
(7,0)       &             0.98        &   0.83   &  0.19 &   0.34  \\
\hline
(10,0)      &      0.85      &  /  &  0.88 &    0.87     \\
\hline
(16,0)      &         0.63    &  /  &  0.61  &   0.57    \\
\hline\hline
\end{tabular}\\
*tight binding parametrization is similar to that in Ref\cite{BLASE}
\end{table}

\end{document}